# The optical properties of transferred graphene and the dielectrics grown on it obtained by ellipsometry

Aarne Kasikov[*], Tauno Kahro, Leonard Matisen, Margus Kodu, Aivar Tarre, Helina Seemen, Harry Alles

*Institute of Physics, University of Tartu, W.Ostwaldi 1, 50411 Tartu, Estonia*

**Abstract.**

Graphene layers grown by chemical vapour deposition (CVD) method and transferred from Cu-foils to the oxidized Si-substrates were investigated by spectroscopic ellipsometry (SE), Raman and X-Ray Photoelectron Spectroscopy (XPS) methods. The optical properties of transferred CVD graphene layers do not always correspond to the ones of the exfoliated graphene due to the contamination from the chemicals used in the transfer process. However, the real thickness and the mean properties of the transferred CVD graphene layers can be found using ellipsometry if a real thickness of the $SiO_2$ layer is taken into account. The pulsed layer deposition (PLD) and atomic layer deposition (ALD) methods were used to grow dielectric layers on the transferred graphene and the obtained structures were characterized using optical methods. The approach demonstrated in this work could be useful for the characterization of various materials grown on graphene.

**Keywords**

Graphene. Dielectrics. Spectroscopic ellipsometry. Optical properties. PMMA.

**Highlights**

- The optical properties of the transferred graphene are found using spectroscopic ellipsometry
- $SiO_2$ layer thickness on Si substrate is obtained together with graphene properties
- The quality of dielectric material on the top of graphene is characterized by optical methods

---

[*] Corresponding author
E-mail: aarnek@ut.ee

**Graphical abstract**

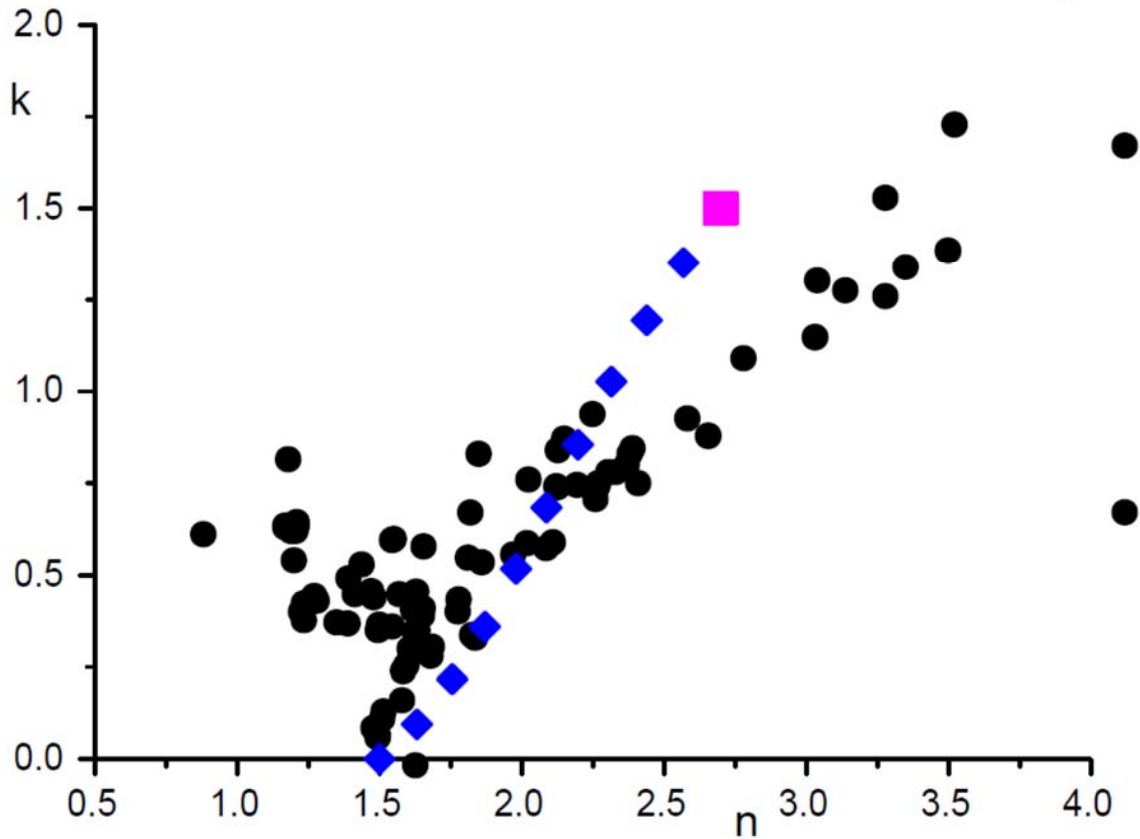

1. **Introduction**

Graphene, a two-dimensional conducting layer of carbon holds great promises for future electronic and optoelectronic applications. These applications often require dielectric layer on top of graphene, but the methods to characterize and control the properties of both graphene and the dielectrics certainly need further development. With micromechanical exfoliation technique the graphene samples with submillimeter size can be extracted [1-3]. In order to get larger graphene samples, the most popular method is currently chemical vapour deposition (CVD) method [4-8]. With this method, graphene is typically grown on Cu-foils at elevated temperatures and then transferred to another, dielectric substrate [9-12].

There is a number of works [2-3, 13-21] dealing with the optical properties of an exfoliated graphene layer. In the work of Weber *et al.* [2] ellipsometric analysis was performed using simultaneous fitting of several spectra taken from monolayer graphene. This resulted in a layer thickness of 0.34 nm which is also a theoretically expected thickness of graphene layer. On the other hand, assuming 2-4 nm thick air-like material between the $SiO_2$ layer and the overlying graphene was needed to obtain accordance with the independently measured thickness of silica in the work of Kravets *et al*. [3]. In the work of Gray *et al*. [13] a thickness of 0.38 nm for the exfoliated graphene on $SiO_2$-coated Si was obtained from the reflection measurements and 0.82 nm from the atomic force microscopy (AFM) measurements. The value of 0.34 nm was used for the graphene layer thickness in the work of Bruna and Borini [14] for reflection measurements. In ellipsometric analysis a value of 0.335 nm or 0.34 nm is usually taken from the distance between atomic planes in graphite [3, 18-21]. In the work of Wurstbauer *et al*. [15] spectroscopic ellipsometry (SE) was used for exfoliated graphene flakes and the best fit for the film thickness was obtained as 0.7 nm from AFM measurements that had to be compared with a theoretical value of 0.34 nm. To solve such a discrepancy, a value of 0.3 nm coming from a possible difference between adhesion of a tip in contact with graphene and $SiO_2$ was introduced as the correction to the film thickness value from AFM measurements [16]. Also in the work of Matković *et al*. [21] an unexpected $SiO_2$ thickness change of 2 nm was obtained over their ellipsometric measurement region and a Cauchy-type layer of water-air mixture was introduced to compensate a difference. In this way, the thicknesses of 0.325 nm for graphene and 0.85 nm for a water layer, correspondingly, were obtained by combining SE and AFM results.

In the work of Ishigami *et al*. [17] a residue originating from a lithography resist was detected that was possible to remove by heating at 400 °C in $Ar/H_2$ mixture and to obtain a thickness of cleaned graphene layer on silica as 0.9 nm in air and 0.42 nm in ultra-high vacuum (UHV) conditions using scanning tunnelling microscopy (STM) measurements. For transferred graphene (TG) on a sapphire substrate, in the work of Matković *et al*. [22] the same graphene thickness value was used for SE, but the film thickness was obtained as 4.55 nm using AFM measurements for the samples where PMMA material from transfer process was removed using acetone, and 0.65 nm for samples after additional annealing at 480 °C in $Ar/H_2$ environment. In this case, some residue islands with the mean thickness of 25 nm remained on the film. As a result, TG has been divided for analysis, as a real graphene and the impurities on it. The same approach was followed in the work of Nelson *et al*. [23] to analyse different properties of multilayer graphene and a buffer layer beneath of graphene layer grown on SiC. Therefore, the dielectric constants of the material remained to depend on the predefined monolayer thickness obtained from multilayer graphitic material.

TG has also been studied by X-Ray Photoelectron Spectroscopy (XPS) [9,24]. In these measurements the layer of poly(methyl methacrylate) (PMMA), used in the transfer process, was removed using acetone. The PMMA residue was retained on graphene after transfer, but it was partly removed using UHV annealing for 3 hours at 300° C [9,10]. Without annealing [24], the "bumps" of PMMA residue were seen on the graphene layer, the bigger islands for higher concentration of the used PMMA solution. In the similar process, the average thickness of the PMMA residue was 1.0 nm according to the AFM measurements in [11]. This thickness was reduced to 0.5 nm after 3 h annealing in 1 mbar $H_2$ atmosphere. At the same time, starting from 350° C temperature, the creation of defects in graphene was seen by Raman measurements.

In order to use graphene for electronic applications, one should to be able to grow dielectric layers on it. This has been done with exfoliated graphene using a sputtered 0.6-nm-thick Al-layer [25], a thin electron-beam evaporated Al- [26], or Ti-layer [27] of thicknesses below 2 nm as a seed layer, which was then covered with an atomic-layer-deposited (ALD) $Al_2O_3$. It is also possible to use e-beam evaporated $Al_2O_3$, some pulses of room-temperature-deposited ALD alumina [28] or dipping CVD-graphene into deionized water [29] to get a seed layer on it and continue with ALD process of alumina at 200 °C. The growth of either amorphous or crystalline $HfO_2$ films by ALD method on exfoliated graphene has been demonstrated with e-beam-evaporated 1.5-nm-thick Al-seed layer [30], with process initiated at 170 °C [31], using $H_2O$ pre-treatment pulses and process initiated at 100 °C [32-33] for both 200 °C $HfO_2$ and $Al_2O_3$, or at 250 °C for $HfO_2$ [12]. For CVD graphene 1-2 nm $HfO_2$ e-beam seed layer has also been used for ALD $HfO_2$ [34]. Smooth MgO films on graphene/SiC substrate have been obtained by reactive pulsed layer deposition (PLD) method [35]. Using sputtering at room temperature and post-oxidation at 250-300 °C, SnO-ZnO films were formed on CVD graphene [36]. On exfoliated BN and $MoS_2$ layers, direct ALD $Al_2O_3$ growth has been shown at 200 °C [37].

The approach using a predefined graphene thickness for exfoliated material does not hinder a comparison of the results obtained by different authors as the possible errors due to the not-exactly-defined graphene thickness influence the results the same way. However, to control technological parameters of chemically produced material it is advisable to use an approach where all the film properties could be found independently. This would also open a way for utilizing optical methods to determine the properties of dielectric materials grown on TG.

## 2. Materials and methods

CVD graphene was grown on 25-µm-thick polycrystalline Cu-foils (99.5%, Alfa Aesar) in a home-assembled hot wall quartz tube reactor. The Cu-foils were annealed for 60 min at 1000 °C in $Ar/H_2$ flow and then exposed to the flow of the mixture of 10% $CH_4$ in Ar (both gases 99.999%, Eesti AGA AS) at the same temperature for 120 min. After growth process the graphene samples were slowly cooled to room temperature in Ar flow. The obtained graphene films on Cu-foils were transferred to oxidized Si substrates with about 300 nm thick thermal oxide ($SiO_2$) layer (MTI Corp) using a wet transfer process. First, a thin layer of PMMA (with the thickness of 100-120 nm, estimated by ellipsometry) was spin coated on top of graphene films on Cu-foils. Unprotected graphene from the other side of Cu-foils was etched off using Ar plasma. Then, PMMA/graphene/Cu structures were floated in aqueous 0.1M $(NH_4)_2S_2O_8$ solution to dissolve copper. PMMA/graphene films were scooped into distilled water to remove the residuals from copper etchant. After that, the films were transferred onto target substrates. Finally, the PMMA layer on graphene was removed using dichloromethane.

The optical measurements of the TG samples were performed on spectroscopic ellipsometer GES-5E (Semilab Co) using a microspot option where light is focussed on a film surface via a telescope. The converging angle of a beam was about 3 deg and the spot size 0.35mm×0.8 mm for 65° or 70° angle of incidence. Fitting was performed using a program WinElli II. Fit quality was characterized using a correlation function between the measured and computed spectra $R^2$ reaching unity for ideal correspondence. Raman spectra of graphene samples were recorded with a Renishaw inVia confocal µ-Raman spectrometer. A 50× objective lens was used to focus the excitation beam of the 514 nm $Ar^+$ laser line to ~1 µm diameter spot on the sample. The accumulation time was set to 10 s and the radiant power was held at less than 1 mW in order to avoid destroying effect on graphene due to local heating. The XPS measurements were carried out with a SCIENTA SES-100 spectrometer using an unmonochromated Mg Kα X-ray source (incident energy = 1253.6 eV), a take-off angle of 90º and a source power of 300 W. The pressure in the analysis chamber was below $10^{-9}$ Torr. The surface morphology of graphene samples was acquired with AutoProbe CP-II (Veeco) AFM microscope using a non-contact mode. Typically, the rms roughness of TG on $SiO_2$/Si substrate varied from 0.3 to 0.8 nm in a 1×1 µm² region.

To obtain the dielectric layers on a TG film, graphene was covered using pulsed layer deposition (PLD) of $ZrO_2$ using a ceramic zirconia pellet as an ablation target. The substrate was held in place by a shadow mask through which $ZrO_2$ was deposited onto graphene. A focused beam of a KrF excimer laser (COMPexPro 205, Coherent; wavelength 248 nm, pulse width 25 ns) was used to deposit a target material. Before starting the PLD procedure, the chamber was evacuated and the substrates were heated in-situ at 150 °C for 1.5 hours and then cooled down to room temperature in order to clean the graphene surface and reduce the effect of traces of contaminants left on graphene. The $ZrO_2$ target was ablated using laser pulse energy density of 2.5 J/cm² in the presence of $5 \times 10^{-2}$ mbar of $O_2$ with other typical process parameters being as follows: laser pulse repetition rate 5 Hz, number of laser pulses 2000, the distance between the substrate and the target 75 mm. The other graphene samples were coated with ALD $Al_2O_3$ film in a home-made reactor [38] using 120 cycles of $Al(CH_3)_3 + H_2O$ process with a cycle consisting of 5 s TMA – 2 s $N_2$ purge – 2 s $H_2O$ – 2 s $N_2$ purge at 100 °C.

## 3. Experiment

The CVD graphene samples grown on Cu-foils were transferred onto Si substrates covered with about 300 nm thick $SiO_2$ layer. The dimensions of the TG films were about 8×8 mm². Both Raman measurements and SE were used as a quality check of the obtained samples. According to Raman characterization, they had a single-layer structure (see Fig. 1). Using ellipsometry, the samples were measured under the incidence angle of 65 or 70 deg. At the first stages of experiment, several measurements for each sample were made – one at the point covered with graphene and in addition 1 or 2 measurements at the edge on a pure Si layer. In this case, the thickness of the $SiO_2$ film was obtained from a measurement over the graphene edge and used as a set parameter in the further analysis for the graphene thickness. The characteristics of the graphene films were obtained using Cauchy dispersion model as

$$n = A + B/\lambda^2 + C/\lambda^4, k = D/\lambda + E/\lambda^3$$

with the graphene thickness as an additional free parameter. The analysis was performed in a 1.3 – 5 eV energy range. In this case, the difference between the model and the real optical parameters at energies above the energy of van Hove peak [3] was sacrificed to get as wide energy region for analysis as possible. In Figure 2 one typical sample of the measurements in points with coated and uncoated graphene is presented. Later it turned out that a better approach is to analyse a spectrum of the measured graphene/$SiO_2$/Si structure as a whole, by setting all the graphene dispersion parameters, as well as the thickness of a $SiO_2$-layer, free. This procedure is successful thanks to a circumstance that a $SiO_2$/Si stack creates the high contrast well-defined ellipsometric spectra, which are further modulated by a thin graphene layer. Then, the fitted dispersion parameters allow us to obtain the modelled refractive index values for every wavelength value. Here, under the term "graphene", we mean the material collected on $SiO_2$ after the transfer process of CVD graphene from Cu-foils.

As the thickness values obtained for TG from spectroscopic ellipsometry were clearly higher than the distance between the graphite atomic planes of 0.335 nm, a composition of the films was checked using XPS. The samples of CVD graphene on Cu-foil, transferred graphene on $SiO_2$/Si before and after the removal of the PMMA layer were analysed. The results are presented in Fig. 3. The XPS results were fitted using Gaussian fit in Origin 8 program, before that the background of the peaks was subtracted using Tougaard model. Due to low depth of origin for XPS signal, the spectrum before PMMA removal corresponds to a pure PMMA material.

For graphene on Cu-foil before the transfer process we see a $sp^2$ C peak at 284.5 eV with lower peaks of $sp^3$, C-O-C and O-C=O probably from air contamination. Then, a sample of TG covered with PMMA layer shows a strong peak of $sp^3$ carbon resulting from PMMA at 284.8 eV with the additional peaks from C-O-C and O-C=O groups. After removing the PMMA layer with dichloromethane, the $sp^2$ and $sp^3$ components of the TG have about the same intensity with other carbon groups being present. This demonstrates that our graphene layer on $SiO_2$, of a single layer nature according to Raman measurements, has a remarkable component of other carbon-containing material. Comparing the XPS results for graphene/Cu films with those of the ellipsometry is not possible due to low contrast of the ellipsometric spectra of copper.

4. Discussion

In order to clear the situation with XPS, we present a set of our results for optical parameters obtained for TG samples from number of batches and using different methods to remove the PMMA layer (Fig.4). Here on the x- and y-axes are the refractive and absorption indices and the points represent the optical parameters of the composite layer of TG at 633 nm wavelength obtained from SE analysis. Each point on the graph corresponds to one transferred graphene sample analysed like presented in previous paragraph. We see the points concentrating around the region $\tilde{n} = (1÷2)-(0.4÷0.8)i$ with a long tail which reaches to the optical parameters high above 2.7-1.5i measured for exfoliated graphene [2-3, 9, 16, 18-20]. The mean graphene layer thickness obtained from SE for this central region is about 2 nm, and about 0.6 nm for the tail part. The higher part of the tail corresponds to our results where the fitted thickness of the graphene layer is abnormally low (a point 2.65-0.9i in Fig. 4 corresponds to a thickness of 0.45 nm). It demonstrates that due to a noise in our measurements, the film thickness should be more than 0.5 nm to be reliably defined within given approach and we cannot check the high-quality layers using only the simplest measurement with our SE apparatus. On the other hand, also for the TG films with the thickness approaching 10 nm the results become unreliable due to low contrast of the refractive index between $SiO_2$ and the graphene film. Still, it is possible to differentiate between the objects with different amount of foreign material, either PMMA or water-air mixture composition, on the TG layer.

As an illustration, we show the possible optical parameters of the film which would consist of the mixture of graphene and PMMA. If we approximate this mixture material with a Bruggeman model

$$\frac{\varepsilon_1-\varepsilon}{\varepsilon_1+2\varepsilon} + (1-f)\frac{\varepsilon_2-\varepsilon}{\varepsilon_2+2\varepsilon} = 0 \,,$$

where $\varepsilon_1 = (2.7-1.5i)^2$, is taken for a pure graphene, and $\varepsilon_2 = 2.25$ for PMMA with the refractive index of 1.5, we can find the dielectric function values for different volume fractions of graphene material from 0.1 to 1. We see that this dependence crosses our set at $f \sim 0.4$-$0.5$ giving us approximately equal volumes of graphene and PMMA in our films. In Fig.5 we see the AFM picture of our graphene layer showing the minor amounts of contamination on the film surface in the local region where there are no cracks in graphene.

The accuracy of ellipsometric analysis depends on contrast of the optical parameters between the sublayers of a film material. Luckily, we are here in a favourite situation due to clearly different optical properties of $SiO_2$ and graphene. It gives us a possibility to differentiate between the spectra taken from the points either covered or not covered with graphene as shown on Fig. 2. For bigger amounts of residue, the effective index of refraction of the residue/graphene film starts to approach a refractive index of the $SiO_2$ layer and the two materials become indiscernible.

Our approach uses information about $SiO_2$ layer and the material after the transfer of CVD graphene obtained at the same point. If the thickness of the silica layer would be measured from the other place near to the graphene layer, the uncertainty on a possible change of the thickness of $SiO_2$ layer would influence the results. This is demonstrated in Fig. 6. The results are presented on a line over a graphene sample 8x8 mm$^2$ with a measurement step of 1 mm. Each measured point was analysed independently. The lines show the results for physical thickness of $SiO_2$ and the summary layer of $SiO_2$ and TG. The upper rectangles present a film thickness in an approach where all the material was treated as $SiO_2$ at fitting. Therefore, the outmost points without TG represent a continuation of the lower (silica thickness) line. The correlation function values for two models for $R^2$ were ~0.95 and ~0.99 for pure $SiO_2$/Si and for TG/$SiO_2$/Si models, accordingly, at the points covered with graphene while for the points over the edge of graphene, $R^2$ value was about ~0.99 for pure $SiO_2$/Si model. It means that though we fit the thicknesses of the two films at the same time, we can make a clear distinction between the two structures. This difference of a $SiO_2$ layer on the Si substrate over the surface may explain the results of Matković et al. [21]. We have come upon the $SiO_2$ layer thickness changes up to 5 nm over a distance of several cm for the commercial wafers.

As we saw in Fig. 2, over most of the spectra, the influence of the graphene material is not big and the main difference we get is in a change of the interference maxima positions due to changing optical thickness of a material deposited on Si. In Figure 7 the results of modelling with a thickness of the TG set free or fixed at 0.3 nm are presented together with the graphene optical parameters from Ref. 3. It is not possible to discriminate between the models with different graphene thicknesses on a base of fit quality here as for both cases, the fit quality $R^2$ equaled to 0.997.

The ellipsometrically characterized samples give us the possibility to get information about the material grown on the graphene. If we have both the thickness of $SiO_2$ layer and the optical parameters for the TG layer, we get a precharacterized substrate and it is possible to use it to find the parameters of a next layer of material over the preformed structure. So we covered the measured and analysed TG film with a layer of PLD $ZrO_2$ and performed the ellipsometric measurements of the material on both Si substrate and on TG. At first, a $ZrO_2$ film on Si was analysed taking into account a thickness of the native oxide on a reference substrate and then, to avoid a pileup of free parameters in the analysis, a film on graphene was fitted with the properties of the film from Si substrate with the additional porosity. The used dispersion models were Tauc-Lorentz having a Lorentz dispersion combined with the absorption edge modified by excitons according to [39], and the $ZrO_2$ dispersion from the experimental values given in WinElli II database. Term "mixed" means here that the material having a pregiven optical dispersion is mixed with a component of voids within it according to Bruggeman inhomogeneity model. The results obtained for $ZrO_2$ are presented in Table I. The spectral region of analysis was 1.3 – 5 eV.

**Table I**

| Film model | Thickness, nm | Porosity | Refractive index at 633 nm | Fit quality $R^2$ |
|---|---|---|---|---|
| Tauc-Lorentz on $SiO_2$/Si | 9.3 | | 1.78-0.007i | 0.991 |
| Mixed $ZrO_2$/Void on $SiO_2$/Si | 9.2 | 0.45 | 1.64 | 0.997 |
| Mixed Tauc-Lorentz on graphene | 11.1 | 0.20 | 1.575-0.005i | 0.996 |
| Mixed $ZrO_2$/Void on graphene | 10.8 | 0.51 | 1.60 | 0.996 |

For Mixed Tauc-Lorentz fit in Table I a porosity is presented relative to the former, Tauc-Lorentz model layer. For mixed $ZrO_2$/Void model, the porosities are given against a dense $ZrO_2$ material and the concentration of the oxide in the deposited material on graphene compared to that on $SiO_2$/Si equals to 0.89. Due to low thickness of the deposited material, the difference in the measured ellipsometric spectra is low and the fits with different used dispersion models result in clearly different refractive indices of the material. Still, the film on graphene is thicker and more porous than that on $SiO_2$/Si substrate in both approximations. The XRF measurement of the $ZrO_2$ layer on graphene yielded the mass thickness of 1.60 µg/cm$^2$ for Zr metal that gives us 10.6 nm thick $ZrO_2$ film for taken porosity of 51%. Scanning electron microscopy (see Fig. 8) demonstrates more structured character of the $ZrO_2$ deposited on TG compared to that on $SiO_2$.

Due to low dispersion of material the ALD $Al_2O_3$ films were analysed using a model of porous alumina only. The results are presented in Table II.

**Table II**

| Film model | Thickness, nm | Porosity | Refractive index at 633 nm | Fit quality $R^2$ |
|---|---|---|---|---|
| Mixed $Al_2O_3$/Void on $SiO_2$/Si | 14.8 | 0.055 | 1.72 | 0.989 |
| Mixed Al$_2$O$_3$/Void on graphene | 16.15 | 0.21 | 1.60 | 0.997 |

Also for ALD process, the film grown on the graphene layer is thicker and more porous than the one grown on Si/SiO$_2$ layer.

In this kind of analysis, one should check that the measurements before and after coating are made at the same point to avoid errors from the SiO$_2$ thickness change. Exactness of the graphene model is of less importance as the graphene is taken into account using its ellipsometric, not physical parameters (see Fig. 7).

## 5. Conclusions.

The CVD graphene layers grown on Cu-foils and transferred to Si substrates with about 300 nm thickness of SiO$_2$ layer using PMMA overcoat retain an amount of carbon residue even after the removal of the PMMA layer. This contamination is manifested in the XPS measurements and must be taken into account in further processing of the transferred graphene layer. Spectroscopic ellipsometry opens a way to characterise the real TG taking into account the residue remaining on the surface of the 0.34 nm thick monolayer graphene film. This optical characterization is facilitated thanks to the silica layer creating a spectral structure with contrast properties due to refractive index changes on the film borders, the background of which makes it easier to mark the differences in spectra despite of the low thickness value of the graphene. In the next step, a way is open to obtain information about the structures grown on the graphene, using ellipsometric measurements. In this way, we have a possibility for optical characterization of the material grown on graphene.

**Acknowledgements**

This work has received funding from the European Union's Horizon 2020 research and innovation programme under grant agreement No 696656 and from Estonian Research Council by institutional grants IUT2-24 and IUT34-27. We wish to thank Jaan Aarik and Aile Tamm for useful discussions.

**Appendix**

As for visual look, so also for ellipsometry SiO$_2$ layer beneath alleviates detecting a graphene layer. The information about a film added is obtained through a modulation of the ellipsometric spectra of the underlying structure like seen on Fig. 2. If we define as $r_{10}$ the Fresnel reflection coefficients for light falling under s- and p-polarization for full structure to which we are going to add a graphene layer, then after graphene transfer we get the Fresnel coefficients $r_{20}$ which can be defined as

$$r_{20} = \frac{r_{21}+r_{10}e^{-2i\delta}}{1+r_{21}r_{10}e^{-2i\partial}} \quad (1).$$

Using this definition, a medium 0 would be all the structure below graphene layer, 1 – graphene layer, 2 – air, $r_{21}$ a Fresnel reflection coefficient for transition air – graphene, but $r_{10}$ would be defined for graphene – substrate structure. Here, $\delta$ is a complex phase thickness of the graphene layer

$$\delta = \frac{2\pi \tilde{n} d}{\lambda} \quad (2).$$

To obtain the reflection conditions involving the pure substrate structural parameters from transition air – substrate, we need to add into the formulae an additional layer of air between the graphene and substrate structure.

In this case, the media 3, 2, 1, 0 would be air, graphene, air, substrate, and full reflection coefficient

$$r_{30} = \frac{r_{32}+r_2 e^{-2i\delta}}{1+r_{32}r_2 e^{-2i\partial}} \quad (3),$$

where $r_2$ is a Fresnel reflection coefficient for the structure graphene - air - substrate. Further,

$$r_2 = \frac{r_{21}+r_{10}}{1+r_{21}r_{10}} \quad (4),$$

due to zero thickness of the air layer. After substitution and taking into account that

$r_{21} = -r_{12} = -r_{32}$, we get

$$r_{30} = \frac{r_{32}(1-r_{32}r_{10})+(r_{10}-r_{32})e^{-2i\delta}}{(1-r_{32}r_{10})+r_{32}(r_{10}-r_{32})e^{-2i\delta}} \quad (5).$$

Now, due to $2i\delta \ll 1$ we can take $e^{-2i\delta} \cong 1 - 2i\delta$, so

$$r_{30} = \frac{r_{10}(1-r_{32}^2)-2i\delta(r_{10}-r_{32})}{(1-r_{32}^2)-2i\delta r_{32}(r_{10}-r_{32})} = \frac{r_{10}(1-r_{32}^2)-2i\delta(r_{10}-r_{32})}{(1-r_{32}^2)\left[1-\frac{2i\delta r_{32}(r_{10}-r_{32})}{(1-r_{32}^2)}\right]} \quad (6).$$

As the last member in denominator is small, so due to $\frac{1}{1-\alpha} \cong 1 + \alpha$ we get

$$r_{30} = r_{10} - \frac{2i\delta(r_{10}-r_{32})}{1-r_{32}^2} + r_{10}r_{32}\frac{2i\delta(r_{10}-r_{32})}{(r_{10}-r_{32})} = r_{10} - \frac{2i\delta(r_{10}-r_{32})}{1-r_{32}^2}(1-r_{10}r_{32}) \quad (7).$$

Therefore, marking the obtained Fresnel coefficient for either of polarizations as $\rho = r - i\delta A$, we have

$$\rho = \frac{R_p}{R_s} = \frac{r_p - i\delta A_p}{r_s - i\delta A_s} \cong \frac{r_p - i\delta A_p}{r_s}\left(1 + \frac{i\delta A_s}{r_s}\right) \cong \frac{r_p}{r_s} - \frac{i\delta}{r_s^2}(r_s A_p - r_p A_s) \quad (8).$$

Here, $r_p$ and $r_s$ are the Fresnel coefficients for p- and s- polarized light components for a substrate without the graphene layer, and

$$A_i = \frac{2(r_i - r_{32i})}{1 - r_{32i}^2}(1 - r_i r_{32i}) \qquad (9),$$

where $r_{32i}$ are the Fresnel coefficients for the interface graphene – air and $i$ denotes a particular polarization. So, the influence of the thin graphene layer on the ellipsometric signal is proportional to its optical thickness while the difference between the coated and uncoated substrate depends on its optical properties as $(r_s A_p - r_p A_s)$.

For reflectance measurements the same possibility to amplify the contrast between the stacks coated and uncoated with the exfoliated graphene flakes has been shown in [40].

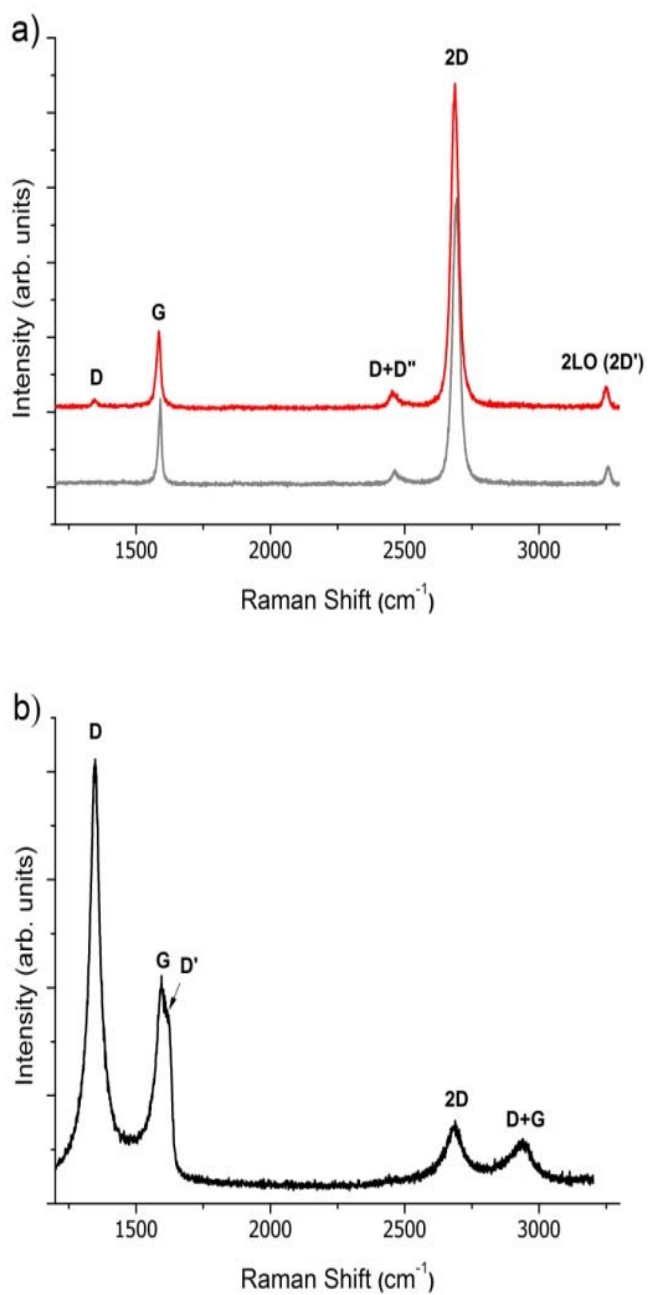

Figure 1. a) Raman spectra of transferred graphene sample on SiO$_2$/Si and graphene after coating with ALD Al$_2$O$_3$ layer; b) Raman spectrum of the graphene after coating with PLD ZrO$_2$ layer. The TMA – H$_2$O ALD process creates only a slight Raman defect band while the graphene is strongly defective after the pulsed layer deposition process.

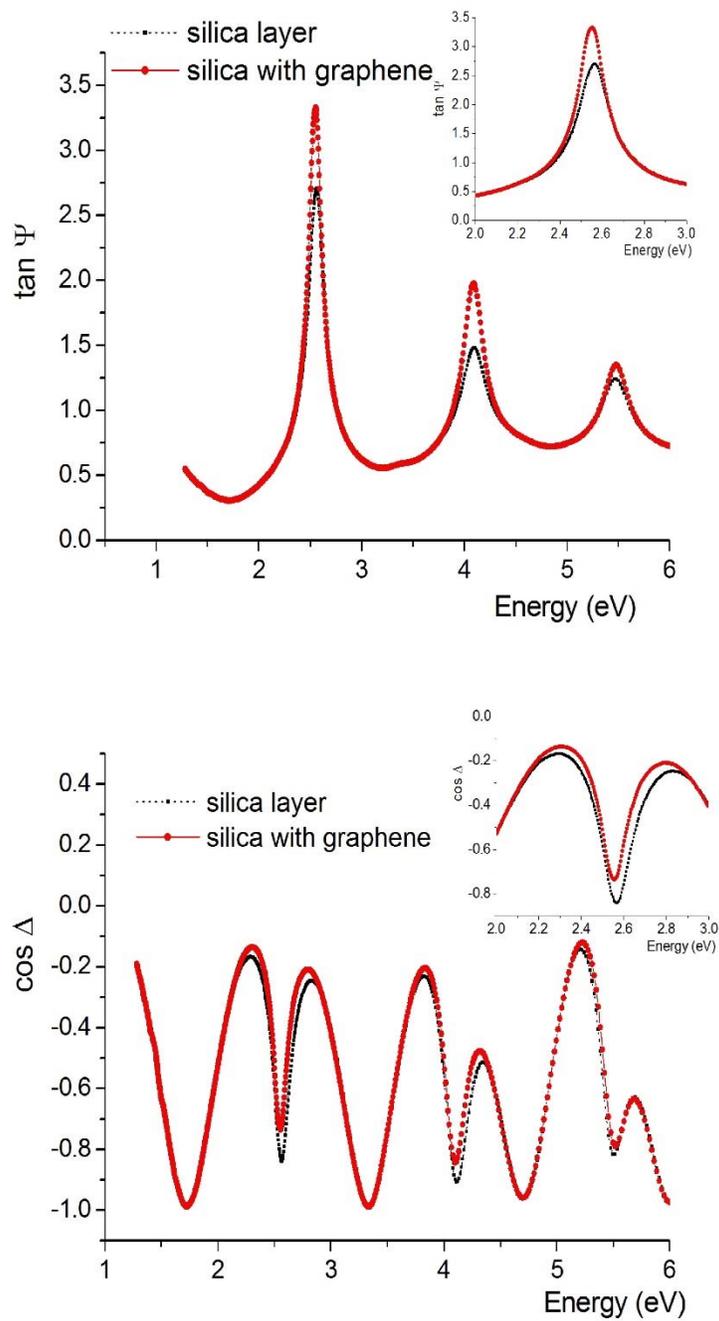

Figure 2. Ellipsometric spectra of the SiO$_2$/Si substrate and of the same substrate with TG.

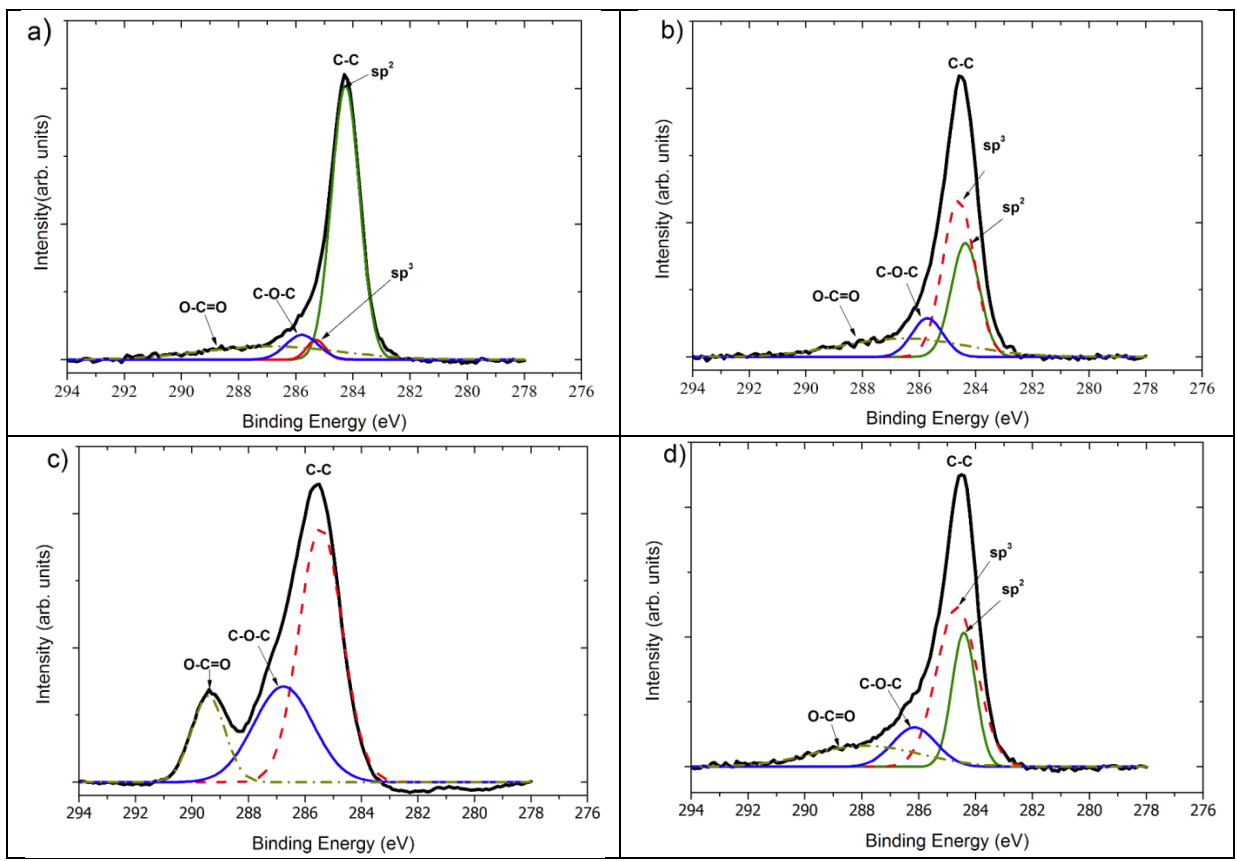

Figure 3. The XPS spectra of graphene C1s line. a,b) slightly and strongly contaminated graphene samples on Cu-foil; c) graphene samples on $SiO_2$/Si substrate before and d) after the removal of the PMMA layer.

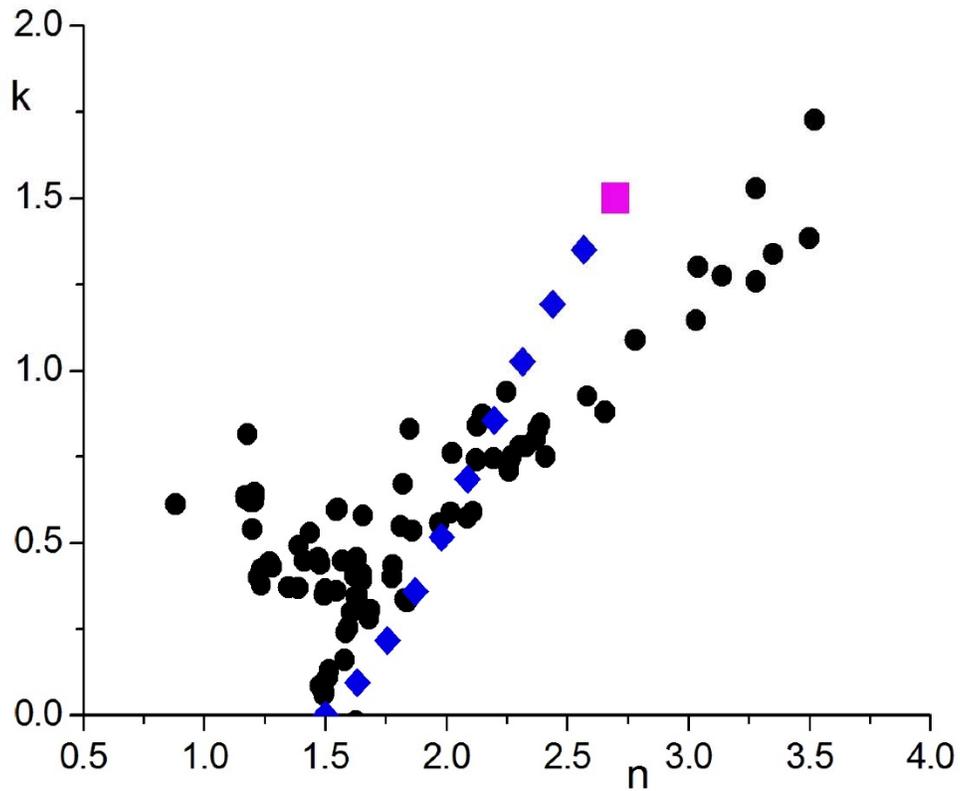

Figure 4. The graphene layer optical parameters (refractive and absorption index) for 633 nm wavelength for different transferred graphene batches. Diagonal rectangles (blue in electronic version) show a Bruggeman mixed layer model of exfoliated graphene and PMMA material for different graphene concentrations with step of 0.1. Big rectangle is a result from [3]. A set of points showing to upper right corner of the figure corresponds to the samples with fitted film thickness less than 1 nm. Each point on Figure corresponds to one TG sample.

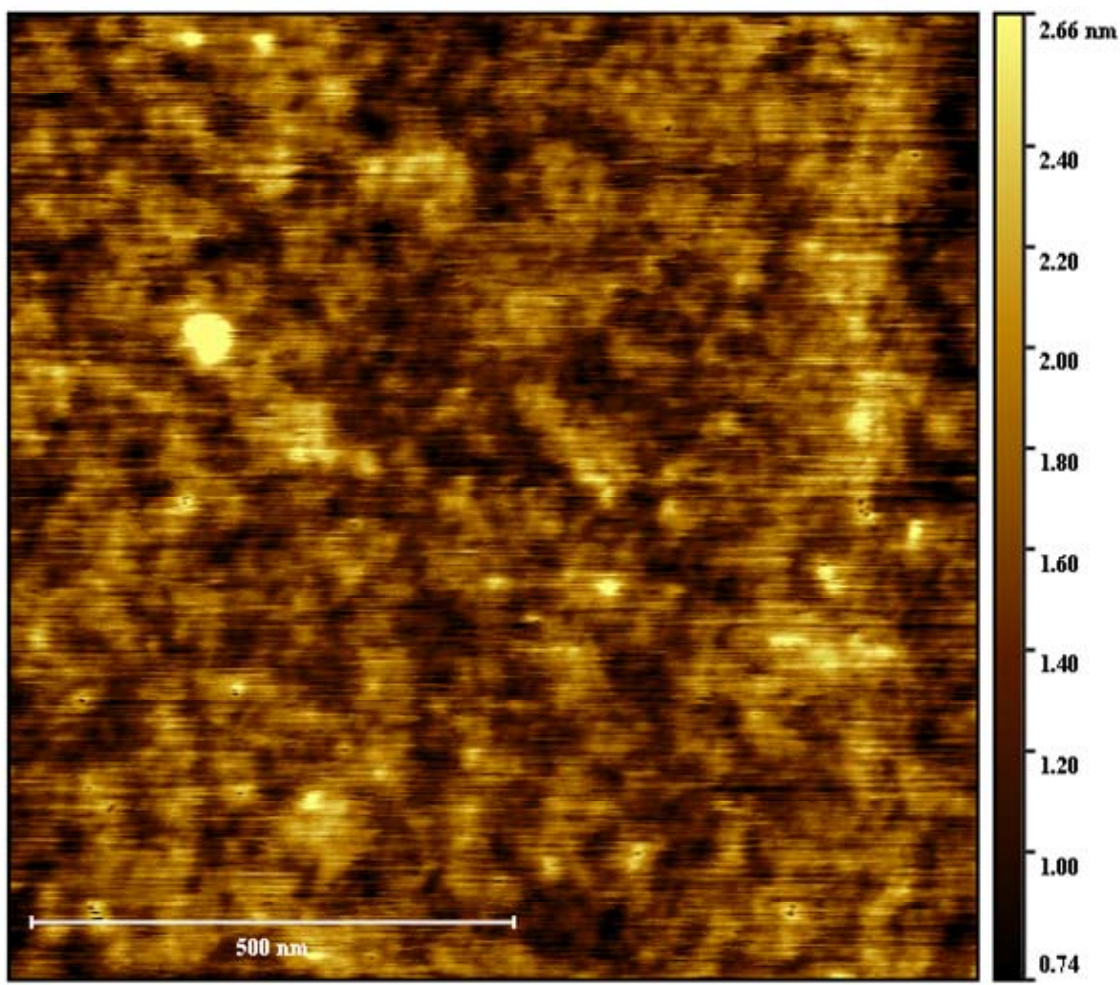

Figure 5. The AFM picture of good-quality TG layer on SiO$_2$/Si substrate. A region without cracks in the graphene layer.

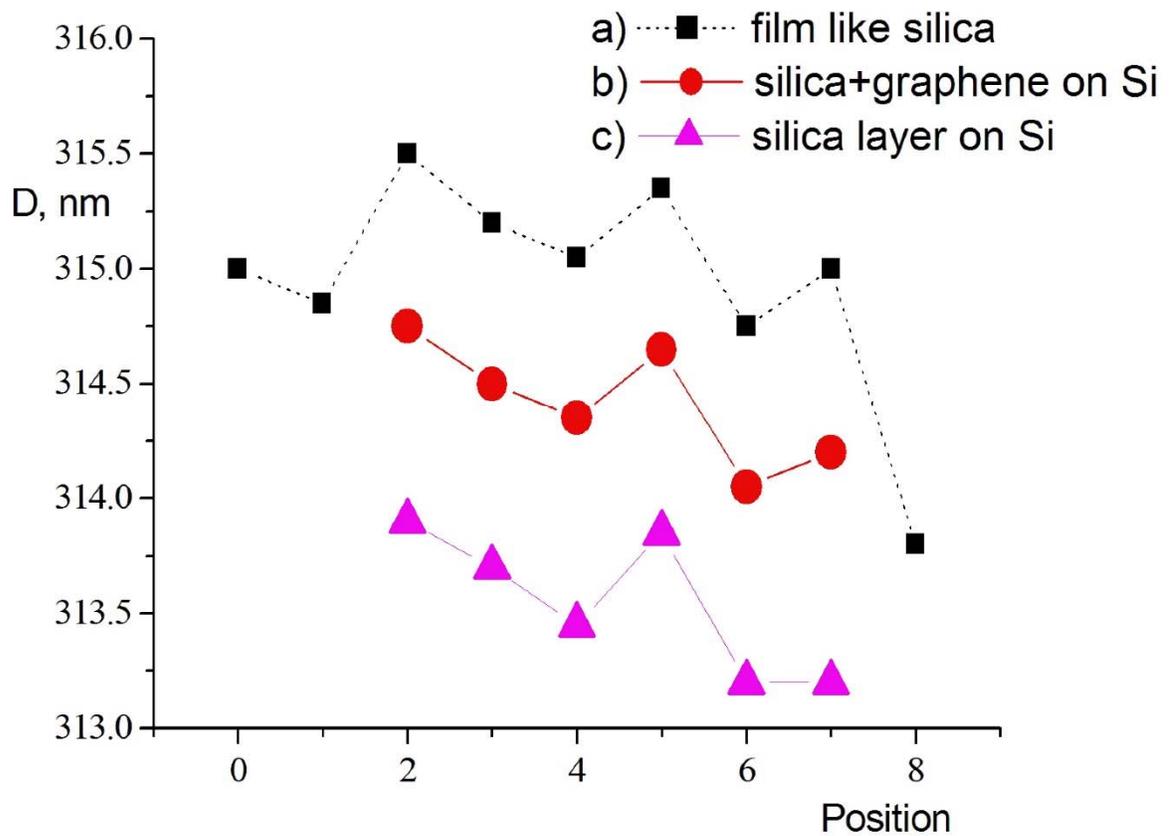

Figure 6. Ellipsometry fitting results of the TG sample on $SiO_2$/Si. a) fitting of the film as pure $SiO_2$; b) film full thickness for a fit as a system consisting of $SiO_2$ and TG layers, graphene situating on $SiO_2$; c) $SiO_2$ thickness in the same points. The graphene sample is on the positions 2 – 7, measurement step 1 mm, mean thickness of the TG layer is 0.85 mm.

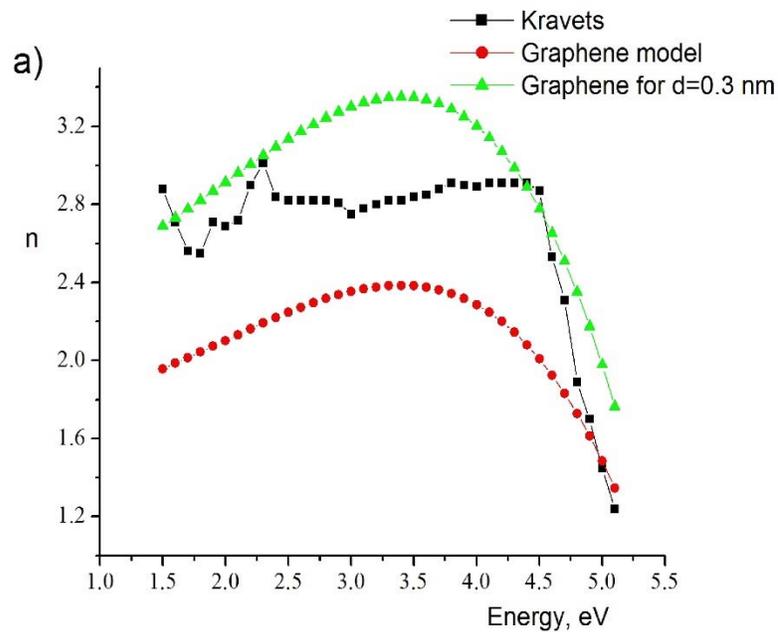

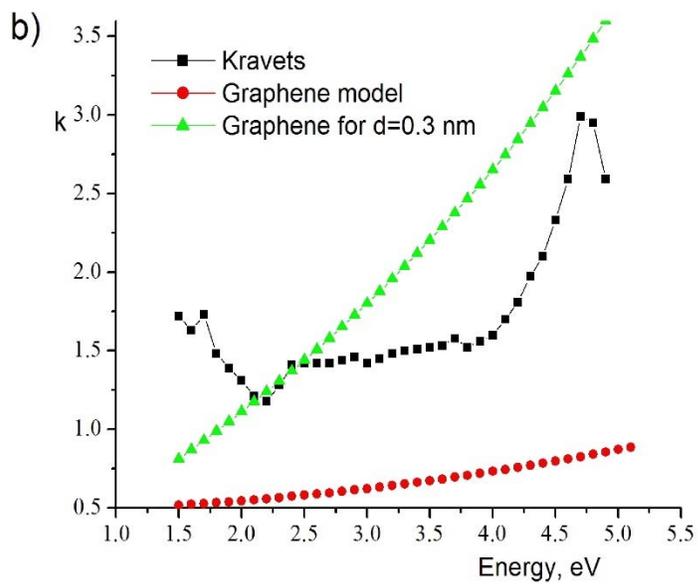

Figure 7. The optical parameters of graphene – exfoliated graphene according to [3], fit of transferred graphene from Fig.6, and fit of TG from Fig.6 if the thickness of the graphene film was fixed at 0.3 nm at fitting. The results are dependent on the used film model.

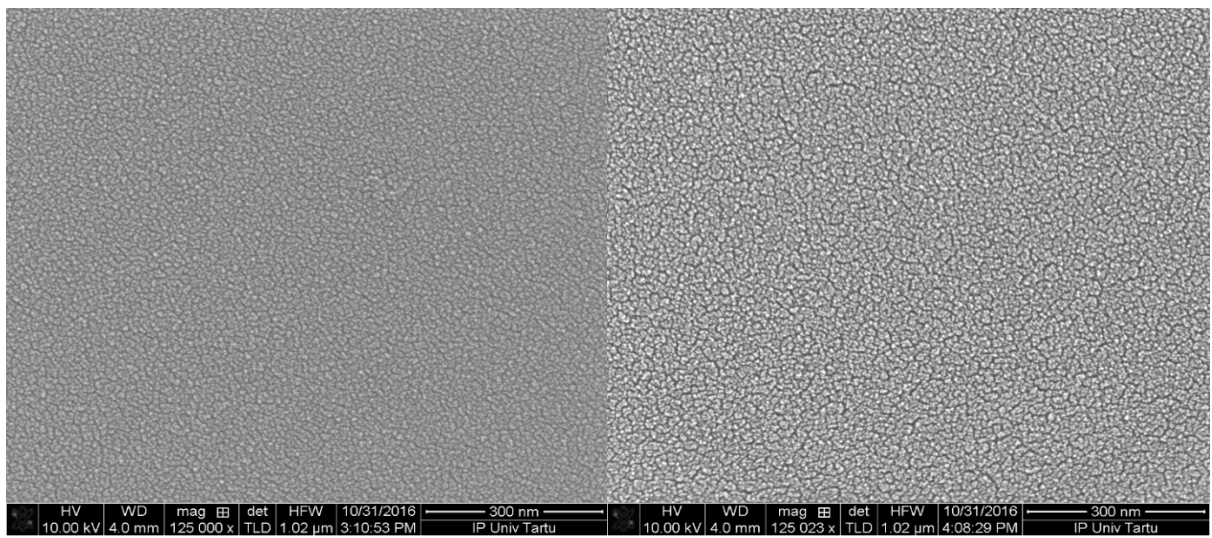

Figure 8. SEM micrograph of the PLD ZrO$_2$ 10 nm film on SiO$_2$/Si (left) and on the TG (right). The Raman spectrum of the sample on TG is presented on Fig.1. The SEM graphs of the ALD Al$_2$O$_3$ samples are not presented as they are indiscernible.